\documentclass[twocolumn]{aastex63}
\usepackage{lineno}
\usepackage{multirow}
\usepackage{amsmath}
\usepackage{array, multirow}
\usepackage{hyperref}
\hypersetup{colorlinks,allcolors=black}
\newcommand{\HI}{H{\sc i}}
\newcommand{\HII}{H{\sc ii}}

\newcommand{\rmHI}{\rm{H\,{\textsc i}}}
\received{....}
\revised{.....}
\accepted{27 February, 2023}
\shorttitle{Oblateness of Dark Matter Halos}
\shortauthors{Das et al.}
\graphicspath{{./}}

\begin{document}
\title{ESTIMATING THE OBLATENESS OF DARK MATTER HALOS USING NEUTRAL HYDROGEN VELOCITY DISPERSION}

\author[0000-0001-8996-6474]{Mousumi Das}
\affiliation{Indian Institute of Astrophysics, 2nd Block Koramangala, Bangalore, Karnataka 560034, India}
\email{mousumi@iiap.res.in, chandaniket@gmail.com}

\author[0000-0003-2476-3072]{Roger Ianjamasimanana}
\affiliation{Instituto de Astrofísica de Andalucía (CSIC), Glorieta de la Astronomía, E-18008 Granada, Spain}

\author[0000-0002-9762-0980]{Stacy~S.~McGaugh}
\affiliation{Department of Astronomy, Case Western Reserve University, 10900 Euclid Avenue, Cleveland, OH 44106, USA}


\author[0000-0003-2022-1911]{James Schombert}
\affiliation{Department of Physics, University of Oregon, 120 Willamette Hall, 1371 E 13th Avenue, Eugene, OR 97403461, USA}

\author{K.~S.~Dwarakanath}
\affiliation{Astronomy and Astrophysics, Raman Research Institute, C.V. Raman Avenue, 5th Cross Road, Sadashivanagar, Bengaluru, Karnataka 560080, India}



\begin{abstract}
We derive the oblateness parameter q of the dark matter halo of a sample of gas rich, face-on disk galaxies. We have assumed that the halos are triaxial in shape but their axes in the disk plane (a and b) are equal, so that $q=c/a$ measures the halo flattening. We have used the \HI~velocity dispersion, derived from the stacked \HI~emission lines and the disk surface density to determine the disk potential and the halo shape at the $R_{25}$ and $1.5R_{25}$ radii. We have applied our model to 20 nearby galaxies, of which 6 are large disk galaxies with M(stellar)$>$10$^{10}$, 8 have moderate stellar masses and 6 are low surface brightness dwarf galaxies. Our most important result is that gas rich galaxies that have M(gas)/M(baryons)$>$0.5 have oblate halos ($q<0.55$), whereas stellar dominated galaxies have a range of $q$ values from 0.21$\pm$0.07 in NGC4190 to 1.27$\pm$0.61 in NGC5194. Our results also suggest a positive correlation between the stellar mass and the halo oblateness $q$, which indicates that galaxies with massive stellar disks have a higher probability of having halos that are spherical or slightly prolate, whereas low mass galaxies have oblate halos ($q<0.55$). 
\end{abstract}




\section{Introduction} \label{sec:intro}
It is now well known for several decades that galaxies are embedded in massive dark matter halos, a clear indication being the non-declining rotation curves of spiral galaxies  \citep{rubin.etal.1980}. However, the detailed properties of the halos themselves -- their masses, density profiles, radii, angular momenta or even shapes are not well constrained. In this study we focus on halo shapes, since recent studies show that it has a strong impact on disk dynamics, such as bar evolution \citep{kumar.etal.2022} or disk warping \citep{debattista.sellwood.1999}. 

Numerical studies have shown that the halo shape evolves in the process of galaxy formation, since mass infall alters halo shapes resulting in rounder and more oblate halos  \citep{dubinski.1994}. Recent cosmological simulations also suggest that galaxy halos are triaxial, with the galaxy plane favouring circular shapes i.e.  b/a$\approx$1 \citep{prada.etal.2019} and the Milky Way type galaxies have spherical or slightly oblate shapes \cite{chua.etal.2019}; twisted halos have also been detected \citep{emami.etal.2021}.  Observational studies have used various methods to constrain halo shape; such as modeling polar rings around galaxies \citep{khoperskov.etal.2014}, flaring of HI disks \citep{becquaert.combes.1997}, gas or stellar kinematics of edge on disks \citep{olling.1996}, globular cluster systems \citep{posti.helmi.2019} and stellar streams \citep{helmi.2020}. Although the studies are few in number it is clear that halo shapes may vary from galaxy to galaxy. The smaller, low luminosity galaxies appear to have oblate halos \citep{hayashi.chiba.2012}, whereas the larger galaxies appear to have all types of halo shapes \citep{peters.etal.2017,banerjee.jog.2008}. The halo shape may also vary within a galaxy as suggested by observations of stellar streams in our Galaxy \citep{bovy.etal.2016} and nearby galaxies \citep{khoperskov.etal.2014}. 

In this paper, we present a new way to determine the halo shapes of galaxies, using the HI velocity dispersion in their extended disks. This is different from previous studies that have used the gas kinematics in the inner disks to model halo shapes \citep{obrien.etal.2010}. Our justification for this approach is that the dark matter halo dominates the mass distribution in the outer disks of galaxies where the stellar disk mass declines. We also use face-on disk galaxies rather than edge-on galaxies, as previously done in the literature \citep{banerjee.jog.2011}. Our basic method is similar to \citet{das.etal.2020}, where we have shown that halo dark matter is important for supporting the vertical structure of extended HI disks. We now take this a step further by modeling the dark matter associated with the disk as part of the halo potential. In the following sections we describe the derivation of an expression for the halo oblateness parameter $q$, and then apply our method to estimate $q$ for a sample galaxies.

\begin{small}
\begin{deluxetable*}{lcccccccccc}
\tabletypesize{\footnotesize}
\tablecaption{Properties of the galaxies\label{tab:galaxy_properties}}
\tablehead{
\colhead{Galaxy} & \colhead{Other} & \colhead{Type} & \colhead{Distance} & \colhead{Spatial Scale} & \colhead{Incl.} & \colhead{R$_{25B}$} & \colhead{R$_{d}$} & \colhead{R$_{d}$} & \colhead{$v_0$} & \colhead{Reference} \\
    & Name &    &  (Mpc) & (pc/\arcsec) & (deg) & (arcsec) & (arcsec) & (Kpc) & \colhead{kms$^{-1}$} & for R$_{d}$   }
\startdata
NGC 0628  & UGC 01149 & SA(s)c, \HII   & 7.3  & 35.4 & 7.0  & 360.0 & 70.7 & 2.5 & 146.3 & \citep{ganda.etal.2009} \\
NGC 3184  & UGC 05557 & SAB(rs)cd      & 11.1 & 53.8 & 16.0 & 255.0 & 56.4 & 0.92 & 158.6 & \citep{tamburro.etal.2008} \\
NGC 4190  & UGC 07232 & Im pec      & 2.83 & 13.7 & 29.0 & 54.0 & 21.1  & 0.29 &  44.3 & \citep{lelli.etal.2016} \\                 
NGC 4214  & UGC 07278 & IAB(s)m        &  2.9 & 14.1 & 44.0 & 330.0 & 62.1 & 0.87 & 76.8 & \citep{hermelo.etal.2013} \\
NGC 4736  & UGC 07996 & (R)SA(r)ab     &  4.7 & 22.8 & 41.0 & 232.9 & 49.8 & 1.1 & 162.2 & \citep{casasola.etal.2017}\\
NGC 5194  &	 UGC 08493;     & SA(s)bc pec    &  8.0 & 38.8 & 41.0 & 232.9 & 78.0 & 3.03 & 209.6 & \citep{casasola.etal.2017}\\
NGC 5236  & UGCA 366  & SAB(s)c        &  4.5 & 21.8 & 24.0 & 465.7 & 54.0 & 1.18 & 200.2 & \citep{bicay.etal.1989}\\
NGC 6946  & UGC 11597  & SAB(rs)cd      &  5.9 & 28.6 & 33.0 & 497.9 & 68.0 & 5.31 & 182.5 & \citep{prieto.etal.2001}\\
Holmberg1 &	DDO 63 & IAB(s)m        &  3.9 & 18.9 &  0.0 & 120.0 & 36.0 & 0.68 & 47.9 & \citep{hunter.etal.2019}\\
Holmberg2 &	DDO50 & Im             &  3.4 & 16.5 & 41.0 & 396.4 & 89.8 & 1.10  & 69.5 & \citep{hunter.etal.2019}\\
M81 dwfA  & PGC 23521 & Irr?       &  3.6 & 17.5 & 23.0 &  37.8 & 14.9 & 0.26  & 32.2 & \citep{hunter.etal.2019}\\
F564-V3   & LSBC D564-08 & .....   &  8.7 & 42.2 & 35.0 &  19.6 & 12.6 & 0.53 & 40.7 & \citep{hunter.etal.2012}\\
IC 10     & UGC 192 & IBm          &  0.7 &  3.4 & 41.0 & 405.0 & 117.8 & 0.40 & 55.8 & \citep{hunter.etal.2012}\\
IC 1613   &  UGC 00668     & IB(s)m         & 0.7  &  3.4 & 37.9 & 303.5 & 170.9 & 0.58 & 51.5 & \citep{hunter.etal.2012}\\
DDO 46    & UGC 3966 & Im          & 6.1  & 29.6 & 28.6 & 66.0  & 38.5 & 1.14  & 49.4 & \citep{hunter.etal.2012}\\
DDO 47    & UGC 3974 & IB(s)m      & 5.2  & 25.2 & 17.0 & 150.0 & 54.4 & 1.37 & 62.9 & \citep{hunter.etal.2012}\\
DDO 53    & UGC 04459 & Im  & 3.6  & 17.5 & 39.5 & 54 & 41.3 & 0.72 & 37.9 & \citep{hunter.etal.2012}\\
DDO 75    & Sextans A & IBm        & 1.3  &  6.3 & 33.5 & 142.8 & 34.9 & 0.22  & 39.8 & \citep{hunter.etal.2012}\\
DDO 187   & UGC 9128  & Im       & 2.2  & 10.7 & 39.0 &  54.0 & 16.9 &  0.18  & 27.3 & \citep{hunter.etal.2012}\\
UGC 8833 & PGC 049452 & Im & 3.1 & 14.9 & 28.0 & 28.8 & 11.3 & 0.17 & 27.3 & \citep{bremnes.etal.1999} 
\\
\enddata
\tablecomments{The disk flat rotation velocity $v_0$ is derived from the baryonic Tully Fsher relation as described in Section~2. The last column gives the reference for the disk scale length $R_d$.}
\end{deluxetable*}
\end{small}

\section{Derivation of halo oblateness factor $q$}

We start with the equation of vertical hydrostatic equilibrium of a disk using \HI~ as a tracer (see \citet{das.etal.2020} for details), where $\sigma_{\rm{zH\,{\textsc i}}}$ is the \HI~ velocity dispersion in the $z$ direction and 
$\rho_{\rmHI}$ is the neutral hydrogen density in the disk. 
\begin{equation}
\frac{1}{\rho_{HI}}\frac{d[{\sigma_{\rm{zH\,{\textsc i}}}}^{2}\rho_{HI}]}{dz}~=~-\frac{d\Phi}{dz}
\end{equation}
The potential is composed of contributions from the stellar disk ($\Phi_{s}$), the \HI ~disk ($\Phi_{g}$) and the halo ($\Phi_{h}$). We will also use the Poissons equation i.e. $\nabla^{2}\Phi=4\pi G\rho$, which in the flat part of disk rotation curve becomes only dependent on the vertical mass distribution.
\begin{equation}
\frac{d^{2}\Phi}{dz^{2}}~=~4\pi G[{\rho}_{s} + {\rho}_{g} + {\rho}_{h}]
\end{equation}
where $\rho_{s}$, $\rho_{g}$, $\rho_{h}$ are the densities corresponding to the stars, gas and halo. As in \citet{das.etal.2020}, we will assume the stellar and gas disks to have exponential forms in the vertical direction, i.e. $\rho_{s}=\rho_{e}e^{-\frac{z}{z_{e}}}$ and  $\rho_{s}=\rho_{e}e^{-\frac{z}{z_{e}}}$, where $\rho_{e}$ and $\rho_{g0}$ are the densities at z=0, and $z_{e}$, $z_{g0}$ are the vertical disk scale lengths for the stellar and gas disks respectively. Let us assume that the halo potential has a logarithmic form, i.e.,
\begin{equation}
\Phi_{h}~=~\frac{1}{2}{v_{0}}^{2}ln(R_{c}^{2} + R^{2} + (z/q)^{2}) 
\end{equation}
where $v_{0}$ is the flat rotation velocity of the disk, $R_{c}$ is the core radius of the halo potential and $q$ is the oblateness of the halo potential or $q=c/a$ and $a=b$. Integrating the first two terms on the right hand side of equation [2] from 0 to z, and replacing the third integral with the derivative of the halo potential $\Phi_{h}$, equation [1] becomes, 
\begin{equation}
\begin{split}
\frac{1}{\rho_{HI}}\frac{d[{\sigma_{\rm{zH\,{\textsc i}}}}^{2}\rho_{HI}]}{dz}~=~-4\pi G [\rho_{e}z_{e}(1-e^{-z/z_{e}}) \\+~\rho_{g0}z_{g0}(1-e^{-z/z_{g0}})]~-~\frac{v_{0}^{2}z}{q^{2}(R_{c}^{2} + R^{2} + (z/q)^{2})} 
\end{split}
\end{equation}
We will take $\rho_{HI}$ from the left hand side onto the right side and integrate throughout from 0 to infinity ($\infty$). Assuming that the density falls to 0 at higher z, and that the mean \HI~disk vertical velocity dispersion is $\sigma_{\rmHI}$, we obtain the following.
\begin{equation}
\begin{split}
\sigma_{\rm{zH\,{\textsc i}}}^{2}\rho_{HI}~=~4\pi G\rho_{e}z_{e}[z_{g0}~-~\frac{z_{e}z_{g0}}{(z_{e}~+~z_{g0})}]\\
~+~2\pi G\rho_{g0}{z_{g0}}^{2}~+~\frac{{v_{0}^2}{z_{g0}^2}}{{q^2}{R^2}}
\end{split}
\end{equation}
For evaluating the last integral we have made an important assumption that,
\begin{equation}
R^{2} \gg \frac{z^2}{q^2}~+~{R_c}^{2}    
\end{equation}
i.e. the radius where the \HI~velocity dispersion is measured, is much larger than the core radius of the halo $R_c$ and the vertical height divided by the halo oblateness $q$. The first part of the assumption is fine as $R_c$ is usually $<<$1 kpc  in galaxies. But the second part of the assumption depends on the value of $q$, as we will see in the following sections. Overall, the assumption hold good as long as equation [5] is applied to large radii in galaxies, where the stellar disk surface density is low. This is typically beyond the $R_{25}$ radius in galaxies.  

For the outer \HI~disks of galaxies at $R>R_{25}$, the stellar mass is barely detected, so that $\rho_{e}\approx 0$, and hence the first term in equation[5] can be dropped. We then obtain the following expression for the halo $q$ in the HI dominated, outer disks of galaxies.
\begin{equation}
q^{2}~=~\frac{{v_{0}}^{2}{z_{g0}}^{2}}{R^2}\frac{1}{({\sigma_{\rm{zH\,{\textsc i}}}}^{2}~-~2\pi G\rho_{g0}{z_{g0}}^2)}   
\end{equation}
An important implication of equation [7] is that for $q$ to be positive, we should have, ${\sigma_{\rm{zH\,{\textsc i}}}}^{2}\gg2\pi G\rho_{g0}{z_{g0}}^2$. But the surface density of \HI~is given by $\Sigma(HI)=2z_{g0}\rho_{g0}$ and the disk dynamical mass surface density by $\Sigma(dyn)=\frac{\sigma_{\rm{zH\,{\textsc i}}}^{2}}{\pi Gz_{g0}}$ \citep{das.etal.2020}. So to obtain an estimate of the halo $q$ in a galaxy from the measurement of the outer disk \HI~velocity, the condition $\Sigma(dyn)>\Sigma(HI)$ should be satisfied. 

In the following sections we will apply equation [7] to determine $q$ for a sample of nearly face-on galaxies. To simplify the equation we will use the fact that the surface density of \HI~is given by $\Sigma(HI)=2{\int_{0}^{\infty}}\rho_{HI}dz=2\rho_{g0}z_{g0}$ \citep{das.etal.2020}. Then equation [7] becomes,
\begin{equation}
q^{2}~=~\frac{{v_{0}}^{2}{z_{g0}}^{2}}{R^2}\frac{1}{({\sigma_{\rm{zH\,{\textsc i}}}}^{2}~-~\pi Gz_{g0}\Sigma(HI)}    
\end{equation}
If the flat rotation velocity $v_0$ is in units of kms$^{-1}$, the vertical disk scale height $z_{g0}$ and the radius in units of kpc, the vertical \HI~velocity dispersion in units of m s$^{-1}$, the gas surface density in $M_{\odot}pc^{-2}$ and the gravitational constant $G=6.67\times10^{-8}$, we obtain,
 
\begin{equation}
q~=~\frac{{v_0}z_{g0}}{R}\times\frac{10^{3}}{[{{\sigma_{\rm{zH\,{\textsc i}}}}^{2}}~-~13.97\times10^{6}z_{g0}\Sigma(HI)]^{1/2}}    
\end{equation}

where $\Sigma(HI)$ includes the correction for helium in the HI gas surface density. In the following sections we will apply the above equation, following the above mentioned units to determine the halo oblateness $q$. 

\begin{small}
\begin{deluxetable*}{lccccccr}
\tabletypesize{\footnotesize}
\tablecaption{Stellar and Gas Masses, and the halo oblateness}
\tablehead{
Galaxy   & M(*)  &    M(HI)    & M(H$_{2}$)     & M(gas)/M(total) & $q$ & $q$ \\
         & (10$^{8}$ M$_{\odot}$) & (10$^{8}$ M$_{\odot}$) & (10$^{8}$ M$_{\odot}$) &   &  ($R_{25}$)  & ($1.5R_{25}$) }         
\startdata
NGC 0628  & 1.50$\times$10$^2$ & 3.80$\times$10$^1$  & 10 & 0.31 &  0.73$\pm$0.36 & 0.74$\pm$0.39  \\
NGC 3184  & 2.34$\times$10$^2$ & 3.07$\times$10$^1$  & 15.85$\times$10$^2$ & 0.22 & 1.15$\pm$0.65 & 1.91$\pm$1.20 \\
NGC 4190 & 1.19 & 0.46  & $<$0.001 & 0.35 &  0.21$\pm$0.07  & 0.11$\pm$0.04  \\
NGC 4214  & 1.06$\times$10$^1$ & 4.08                & 10$^{7}$    & 0.35   & 0.46$\pm$0.18 & 0.36$\pm$0.14 \\
NGC 4736  & 3.15$\times$10$^2$ & 4.0                 & 10$^{8.6}$  & 0.03  & 0.83$\pm$0.31 & 0.69$\pm$0.26 \\
NGC 5194  & 8.38$\times$10$^2$ & 2.54$\times$10$^1$  & 10$^{9.4}$  & 0.08  & 1.27$\pm$0.61 & 0.79$\pm$0.30 \\
NGC 5236  & 6.88$\times$10$^2$ & 1.70$\times$10$^1$  & 3.2$\times$10$^1$ & 0.09  & 0.44$\pm$0.16 & 0.40$\pm$0.15 \\
NGC 6946  & 4.10$\times$10$^2$ & 4.15$\times$10$^1$  & 10$^{9.6}$  & 0.21  & 0.97$\pm$0.53 & 1.50$\pm$0.58 \\
Holmberg1 & 0.37               & 1.55                & $<$0.17     &  0.85  & 0.20$\pm$0.07 & 0.21$\pm$0.07 \\
Holmberg2 & 2.89               & 5.95                & $<$0.40     &  0.74   & 0.57$\pm$0.26 & 0.79$\pm$0.32 \\
M81 dwfA  & 0.16               & 0.25                & 0           &  0.68 & 0.27$\pm$0.08 & 0.16$\pm$0.06 \\
F564-V3   & 0.73               & 0.41                & $<$0.01    &   0.44 & 0.52$\pm$0.11 & 0.32$\pm$0.11 \\
IC 10     & 3.72               & 0.60                & 0.02        &  0.19  & 0.21$\pm$0.07 & 0.17$\pm$0.06 \\   
IC 1613   & 2.79               & 0.34                & 0.05       &  0.16      & 0.91$\pm$0.34 & 0.67$\pm$0.25  \\
DDO 46    & 0.26               & 1.86                & $<$0.06    & 0.91       & 0.53$\pm$0.21 & 0.39$\pm$0.15  \\
DDO 47    & 2.07   & 3.89   & ....    & 0.72       & 0.41$\pm$0.14 & 0.28$\pm$0.10  \\
DDO 53    & 0.25 & 0.52 & ...   & 0.74 & 0.31$\pm$0.11 & 0.27$\pm$0.09  \\
DDO 75    & 0.20    & 0.57      & $<$0.02    &  0.84 & 0.20$\pm$0.07 & 0.11$\pm$0.04   \\
DDO 187   & 0.08               & 0.13                & $<$0.004   & 0.69  & 0.12$\pm$0.04 & 0.09$\pm$0.03\\
UGC 8833  & 0.08 & 0.13 & ...  & 0.68 & 0.19$\pm$0.05 & 0.12$\pm$0.03
\enddata
\end{deluxetable*}
\end{small}

\section{Sample Galaxies and the Data}
 Our model can include only quiescent disks, and hence should satisfy the criteria described in \citet{das.etal.2020}, which are, (i)~the galaxies should have extended \HI~disks that show very little star formation and hence are in hydrostatic equilbrium; (ii)~the galaxies should be close to face-on and (iii)~they should be nearby, so that several values of the azimuthally averaged \HI~velocity dispersion $\sigma_{\rm{zH\,{\textsc i}}}$ over the disk can be obtained. This last point is important for obtaining a mean value of $q$ over the disks. There were several nearly face-on galaxies that we had to exclude because they are tidally interacting and have disturbed morphologies.  

The galaxies in our sample were taken from the {\it The H\,{\sc i} Nearby Galaxy Survey ({\it THINGS})} \citep{walter.etal.2008}, which includes mainly large gas rich galaxies and the {\it Local Irregulars That Trace Luminosity Extremes; The H\,{\sc i} Nearby Galaxy Survey} ({\it LITTLE THINGS}) \citep{hunter.etal.2012} which includes gas rich irregular and dwarf galaxies. We also included two galaxies NGC 4190 and UGC 8833 from the VLA ACS Nearby Galaxy Survey Treasury (VLA ANGST) survey \citep{ott.etal.2012}. Table~1 shows the sample of 20 galaxies, of which first 8  are relatively large, massive spirals (NGC 0628, NGC 3184, NGC 4736, NGC 5194, NGC 5236 and NGC 6946), and the remaining 12 are smaller spiral, dwarf or irregular galaxies. 

The procedure for determining the \HI~ velocity dispersion across the galaxy disks as a function of radius is described in \citet{ianjamasimanana12} and \citet{ianjamasimanana.etal.2017} as well as in \citet{das.etal.2020}. We basically co-added the individual velocity profiles over radial bins to obtain azimuthally averaged, high signal-to-noise ratio (S/N) velocity dispersion values for each radial bin. The stacked profiles were fitted with single Gaussian functions, where the half width of the fitted Gaussian represents the velocity dispersion of the \HI~ gas. This method gives a better estimate of the velocity dispersion compared to the moment2 maps. Figure~4 in the appendix shows the radial variation of the velocity dispersion across the galaxy disks. Its value is less than 10 kms$^{-1}$ at the R$_{25}$ radius, except for a few cases (NGC 5194, NGC 4190, DDO 187 and IC 10), where $\sigma_{\rm{H\,{\textsc i}}}$ lies between 11 and 14 kms$^{-1}$. 

The stellar masses of the galaxies were derived from mid-infrared (MIR, 3.6 $\mu m$) Spitzer IRAC images of the Spitzer Infrared Nearby Galaxies Survey \citep{kennicutt.etal.2003}. For uniformity we have used a mass to light ratio of 0.5 for all the galaxies \citep{mcgaugh.schombert.2014}, although for the dwarfs $M/L=0.4$ is often more appropriate \citep{schombert.etal.2022}. The stellar masses are shown in Table~1. The stellar surface mass densities for 6 of the galaxies in Table~1 are shown in \citep{das.etal.2020}, and are similar to the others in our sample. The parameters required for $q$ determination are $v_{0}$, $z_{g0}$, $\Sigma(HI)$, and $\sigma_{\rm{zH\,{\textsc i}}}$. Since the galaxies are close to face-on, we used the baryonic Tully-Fisher relation \citep{mcgaugh.2012} to derive disk rotation velocities $v_{0}$, i.e. 
\begin{equation}
v_{0}~=~[\frac{M(*)+1.36M(HI)+1.36M(H_{2})}{47}]^{\frac{1}{4}}    
\end{equation}\\
where M(*) is the galaxy stellar mass, M(HI) is the \HI~mass and M($H_2$) is the molecular hydrogen ($H_2$) gas mass (Table~2). The 1.36 factor is due to the presence of helium \citep{leroy.etal.2008,asplund.etal.2009} The galaxy distances have been adopted from the \HI~ surveys. The disk vertical scale length $z_{g0}$ has been derived from the disk scale radius $R_d$ using the empirical relation $\frac{R_d}{z_{g0}}=8.5\pm2.9$ \citep{kregel.etal.2002}. Table~2 also shows the different studies from which $R_d$ has been obtained. The \HI~velocity dispersion is corrected for inclination using the formula mentioned in \citet{das.etal.2020}.  

\section{Results}
In our model, $q$ has been defined assuming the hydrostatic and vertical equilibrium of an \HI~disk. Hence, $q$ can be best determined only in the outer disks of galaxies, where the contribution of the star formation associated with the stellar disk is very low. So we have determined $q$ at two radii, $R_{25}$ and $1.5R_{25}$. For these radii the stellar disk contribution is extremely low and can be neglected (see Figures 3 and 4 in \citet{das.etal.2020}).  

\begin{figure*}
    \begin{centering}
         \includegraphics[width=18cm]{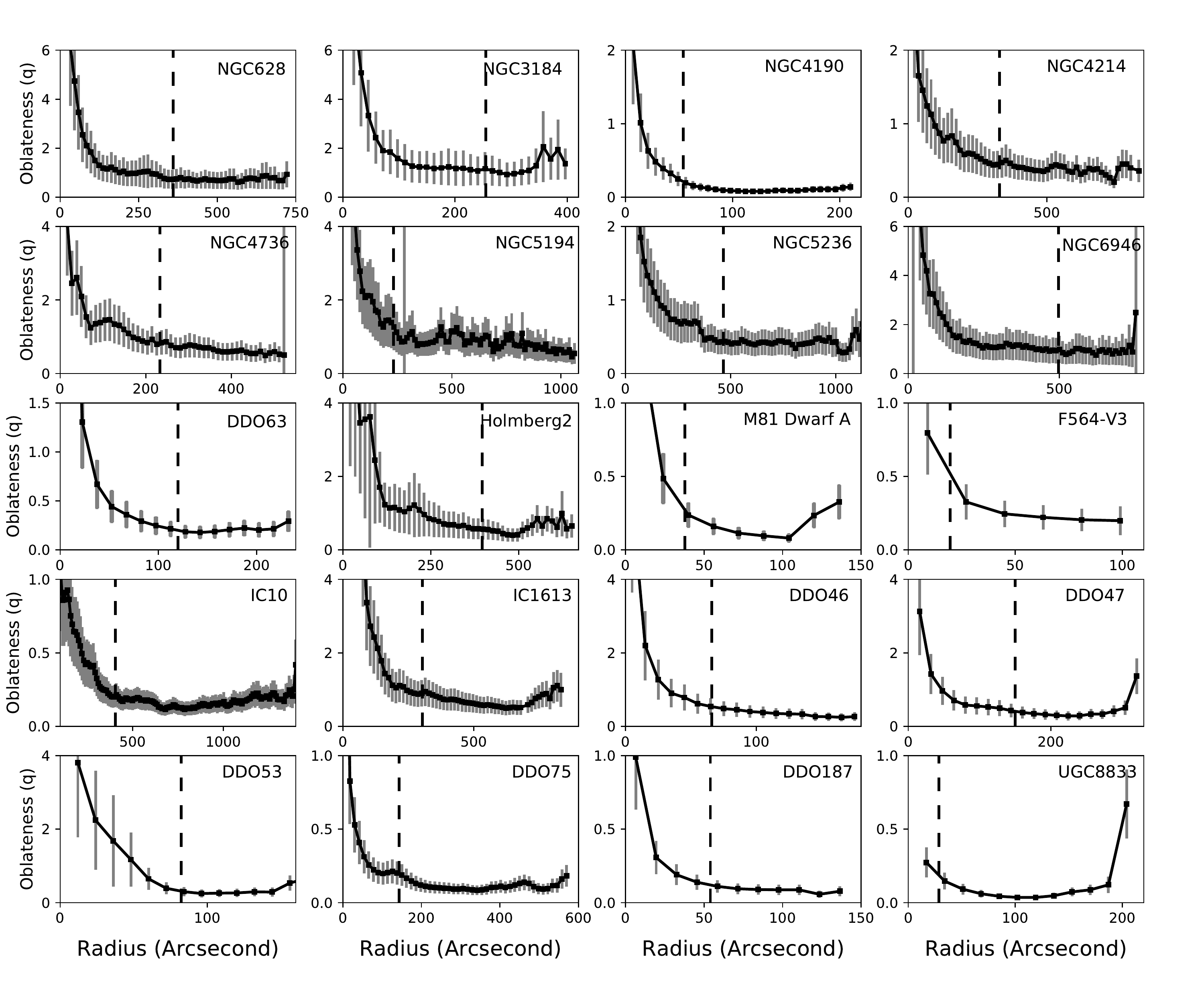}
         \hspace{-2cm}
    \end{centering}     
    \caption{The halo oblateness parameter $q$ as measured from the \HI~velocity dispersion plotted against the radius of the galaxy, for the large galaxies that have stellar masses $M(*)>10^{9} M_{\odot}$ (first two rows) and for the dwarf galaxies that have stellar masses $M(*)<10^{9} M_{\odot}$. The dashed line is the $R_{25}$ radius.}
    \label{fig:area}
 \end{figure*}

\begin{figure}
  \hspace{-0.5cm}
    \begin{centering}
\includegraphics[width=8.5cm]{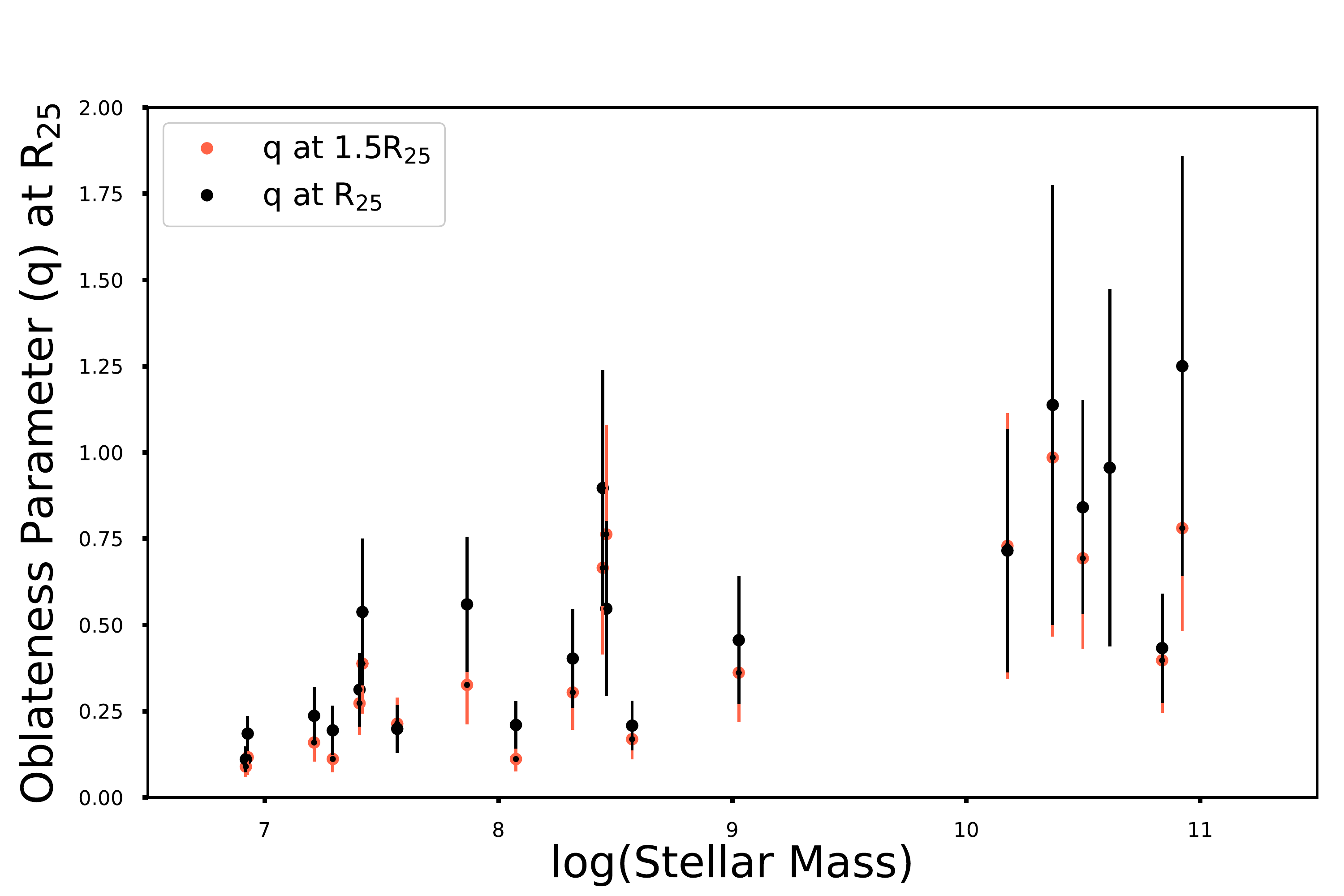}
    \end{centering}     
    \caption{The halo oblateness parameter $q$ measured at the $R_{25}$ (black) and $1.5R_{25}$ radii (red) of the sample galaxies, plotted against stellar mass. }
 \end{figure}

\subsection{The radial variation of $q$ in the disk} 
A clear trend is that $q$ does not vary significantly beyond the $R_{25}$ radius (Figure 1). For the 6 large galaxies that have $M(*)>10^{9}M_{\odot}$, $q$ is close to constant in the outer disk region, except for NGC 3184 which shows a spike in $q$ at large radii. However, this spike maybe due to errors in $\sigma_{\rm{H\,{\textsc i}}}$ measurement as the \HI~column density decreases at the disk edge. A similar trend is present for the less massive galaxies ($M(*)<10^{9}M_{\odot}$), where M81dfwA shows a similar spike in $q$ at the disk edge. The dwarf galaxy IC1613 definitely has q$\sim$1 around $R_{25}$ and 0.5$<$q$<$1.0 beyond that. This is an exceptional case for the dwarfs. The gas rich dwarf DDO46 is the only galaxy that shows a significant decrease in $q$ from $q=0.5$ at $R_{25}$ to $0.3$ at $2R_{25}$, which indicates a fairly rapid change in halo shape with increasing radius. 

\subsection{Variation of $q$ with galaxy stellar mass}
Figure~2 and Table~2 show that galaxies with more massive stellar masses (M(*)$>$10$^9$M$_{\odot}$) have $q$ values close to $q\sim 1$ at $R_{25}$ and beyond (e.g. NGC 628 and NGC 6946). The exceptions are NGC 4214 and NGC 5236. Of these two galaxies, NGC 4214 is a relatively small galaxy and close to being a dwarf \citep{olling.1996}. In fact, based on its rotation velocity, stellar mass, and absolute B-band magnitude values, NGC 4214 has been classified as a dwarf galaxy by \citet{2008AJ....136.2782L}. But NGC 5236 is quite massive ($M(*)\sim 6.9\times10^{10}M_{\odot}$ and so its low $q$ value is quite surprising. For the less massive galaxies, a large fraction of which are LSB dwarfs, $q$ is much smaller (Figure~2 and table~2). Six of them have $q=0.2-0.3$ (Holmberg1, M81dwfA, IC10, DDO75, DDO187 and UGC8833) and 3 of them have $q=0.4-0.5$ (Holmberg2, F564-V3, DDO46). The exception is IC1613 which has a relatively large $q\sim1$ value at $R_{25}$. Thus although 14 of the 20 galaxies follow a consistent trend of larger $q$ with increasing stellar mass, there are clearly exceptions.  

Figure~2 also shows that there is a trend for less massive galaxies to have more oblate halos compared to massive galaxies. However, there is significant error which mainly arises from the uncertainty in disk thickness and from $\sigma_{\rm{H\,{\textsc i}}}$ measurements. Also, the overall $q$ decreases slightly but significantly from the $R_{25}$ radius to the outer $1.5R_{25}$ radius indicating  that halos generally become flatter as the radius increases. We calculated the weighted correlation coefficient using the {\it wCorr} program in the R Statistical Software package (v4.1.2; R Core Team 2021). We found that at the $R_{25}$ radius the Pearson weighted  correlation is 0.78 and the Spearman weighted correlation is 0.79, which are both significant. At larger radii of $1.5R_{25}$, the values are lower, at 0.69 and 0.71 respectively.  

\subsection{Variation of $q$ with stellar and gas mass}
The more massive galaxies have lower gas mass fractions compared to the gas rich dwarf galaxies (Table~1). To see if the gas mass fraction has any correlation with $q$ we plotted the ratio of gas to baryon mass (Figure~3), where,
\begin{equation}
M(baryon)=M(*)+1.36M(HI)+1.36M(H_{2}) 
\end{equation}
so that $M(gas)=1.36M(HI)+1.36M(H_{2})$. This means that the the gas rich galaxies have M(gas)/M(baryon)$>$0.5 and lie at the right hand side, whereas the stellar dominated disks lie towards the left side of the plot. It is interesting to note that the gas rich galaxies (9 of them) all have oblate halos, with $q<0.6$ but the stellar dominated galaxies have a range of $q$ values with $q\approx0.2$ (very oblate) to $q\approx1.25$ (prolate).

\section{Implications of our results}
Perhaps the most important result from this study is that the gas rich galaxies all have oblate halos at $R_{25}$ (Fig. 3). There are 9 gas rich galaxies that have $M(gas)/M(baryon>0.5$ and they are all dwarf galaxies : Holmberg1, Holmberg2, M81dwfA, DDO46, DDO53, DDO75, DDO187, F564-V3 and UGC8833. The $q$ varies between 0.11 in DDO187 to 0.55 in Holmberg2. The stellar dominated galaxies, however, have a variety of $q$ values ranging from $q=0.21$ in IC10 to $q=1.25$ for NGC5194 at $R_{25}$. This is at first surprising as one would expect that a larger stellar mass would make the disk potential more concentrated around $z=0$ plane, which would  produce a larger gravitational force on the halo towards the disk plane. This is similar to the adiabatic contraction of halos during galaxy formation epochs \citep{gnedin.etal.2004}, but instead we are seeing the opposite trend. 

The explanation could be due to 2 factors. (i)~A higher halo spin in the gas dominated galaxies which are all low mass dwarfs. The higher halo spin is due to tidal torques experienced during galaxy formation epochs \citep{peebles.etal.1969}. This also results in the spreading of the stellar disk, leading to lower disk surface densities and hence  lower star formation rates \citep{kim.lee.2013}. So Figure 3 maybe indirect evidence that gas rich dwarfs have larger halo spins, as well as more oblate halos compared to stellar dominated, massive galaxies. (ii)~Secondly a larger concentration of baryons can affect halo particle orbits, resulting in rounder halos, especially in the inner disks of galaxies \citep{cataldi.etal.2021}. Galaxy winds and outflows associated with nuclear activity may also play an important role in making halos rounder \citep{chua.etal.2021} and this may ultimately affect galaxy morphology. Cosmological simulations show that galaxy mergers affect the disks of massive galaxies but affect the halo spin of only dwarf galaxies \citep{rodriguez-gomez.etal.2017}. These results are similar to what we see in our study. Studies also show that halo shapes can vary with radius \citep{emami.etal.2021}. On the observational side recent {\it GAIA} observations indicate that our Galaxy halo shape is not prolate, but instead closer spherical in shape \citep{hattori.etal.2021}. Assuming that our Galaxy has a stellar mass of $M(*)=10^{11} M_{\odot}$, Figure 4 indicates that the $q$ for the Milky Way halo is close $q=1$ (i.e. spherical) at a radius of $R=1.5R_{25}$. The errors have been derived using the propagation of errors formula. Note that the errors in $q$ are a bit high. This is due to the large error bar in the disk height and radius ($z-R_{d}$) scaling relation. However, the intrinsic distribution of disk thicknesses is large. So it is an irreducible uncertainty because for a given face-on galaxy we do not know if it is of median thickness, or somewhat thin, or fat. It can be reduced by using a larger sample of galaxies to derive a correlation between stellar mass M(*) and $q$, which will be useful for constraining cosmological models of galaxy formation and evolution. 

\begin{figure}
\hspace{-0.2cm}
\centering
         \includegraphics[width=8.5cm]{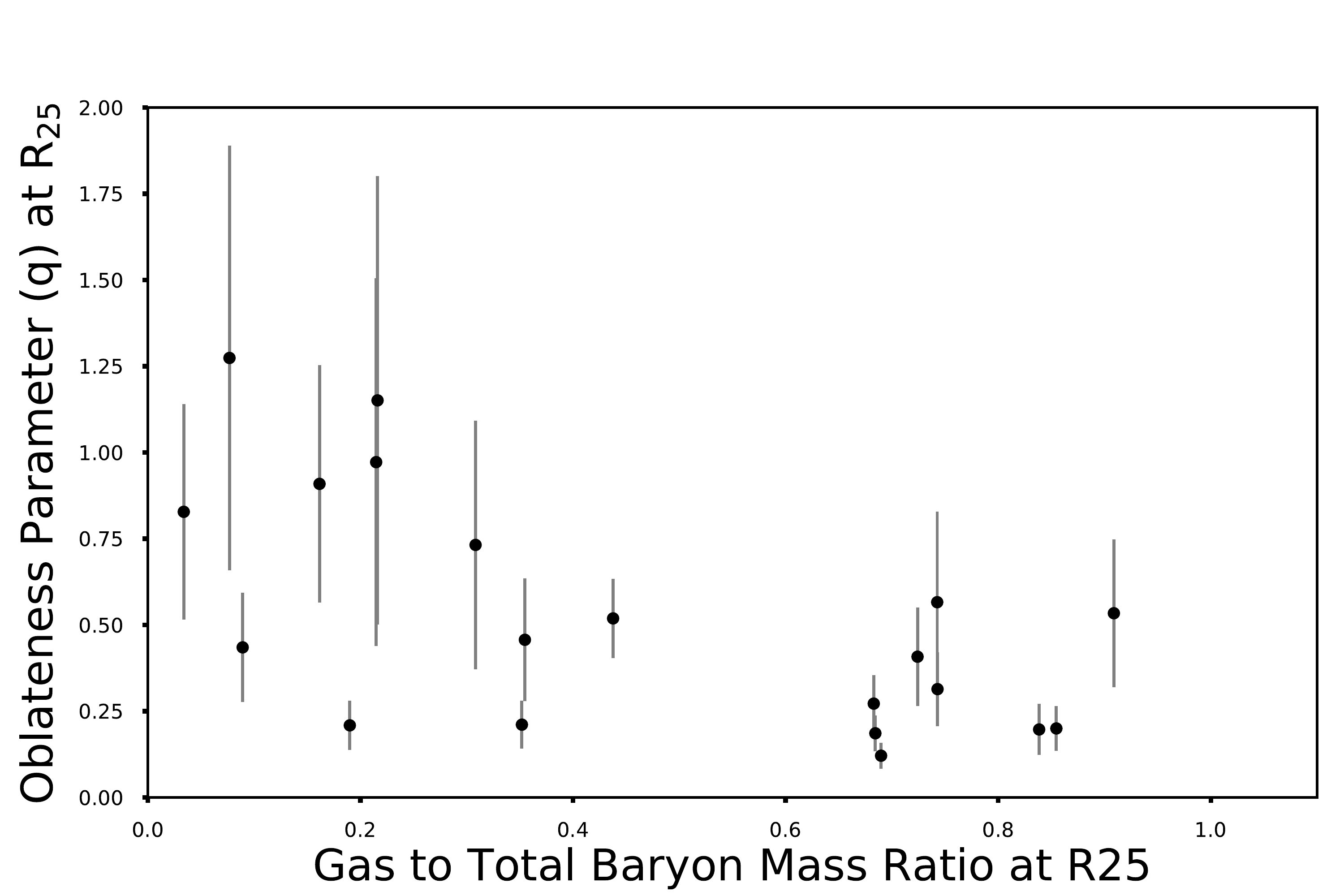}
    \caption{Left~:~The halo oblateness parameter $q$ measured at the $R_(25)$ radius of a galaxy plotted against gas mass to total disk mass or baryon ratio, where the baryon mass is given as M(baryon)=M(*)+1.36M(HI)+1.36M(H$_2$), where the 1.36 factor is the correction for the presence of helium. The gas mass is thus M(gas)=1.36M(HI)+1.36M(H$_2$) }
 \end{figure}

\section{Conclusions}
We have derived the halo oblateness parameter q of a sample of 20 gas rich, face-on disk galaxies using the \HI~velocity dispersion of their disks. Of these galaxies, 6 are large spirals with stellar masses $>10^{10}M_{\odot}$ and the remaining have smaller stellar masses. We find that the q remains fairly constant in the outer disk regions beyond the $R_{25}$ radius. Our study suggests that there is a significant correlation between the stellar mass and the halo $q$ parameter where the weighted correlation coefficient is $\sim$0.78. This  indicates that galaxies with massive stellar disks have a higher probability of having spherical or slightly prolate halos ($0.2<q<1.3$) whereas the low mass dwarfs have oblate halos. On comparing $q$ with the gas mass fractions, we find that the gas rich galaxies with $M(gas)/M(baryons)>0.5$ all have oblate halos ($q<0.55$), whereas stellar dominated galaxies have a range of $q$ values. Since the gas dominated ones are all dwarfs, we conclude that gas rich dwarf galaxies have oblate halos whereas the larger galaxies have a range $q$ values. 

\vspace{-5mm}

\acknowledgements
M.D. acknowledges the support of the Science and Engineering Research Board (SERB) MATRICS grant MTR/2020/000266 for this research. RI acknowledges financial support from the grant CEX2021-001131-S funded by MCIN/AEI/ 10.13039/501100011033, from the grant IAA4SKA (Ref. R18-RT-3082) from the Economic Transformation, Industry, Knowledge and Universities Council of the Regional Government of Andalusia and the European Regional Development Fund from the European Union and financial support from the grant PID2021-123930OB-C21 funded by MCIN/AEI/10.13039/501100011033, by "ERDF A way of making Europe" and by the "European Union" and the Spanish Prototype of an SRC (SPSRC) service and support funded by the Spanish Ministry of Science and Innovation (MCIN), by the Regional Government of Andalusia and by the European Regional Development Fund (ERDF).
\vspace{5mm}

\facilities{VLA}
\software{AIPS}, 
Astropy \citep{astropy2018}, NumPy \citep{numpy2020}, R Statistical Software (v4.1.2; R Core Team 2021) .
\bibliography{sample63}{}
\bibliographystyle{aasjournal}

\appendix
\section{Appendix : The \HI~velocity dispersion Plots}
\label{appendix:a}
\begin{figure*}
    \begin{centering}
        \hspace{0.25cm}
     \includegraphics[width=18cm]{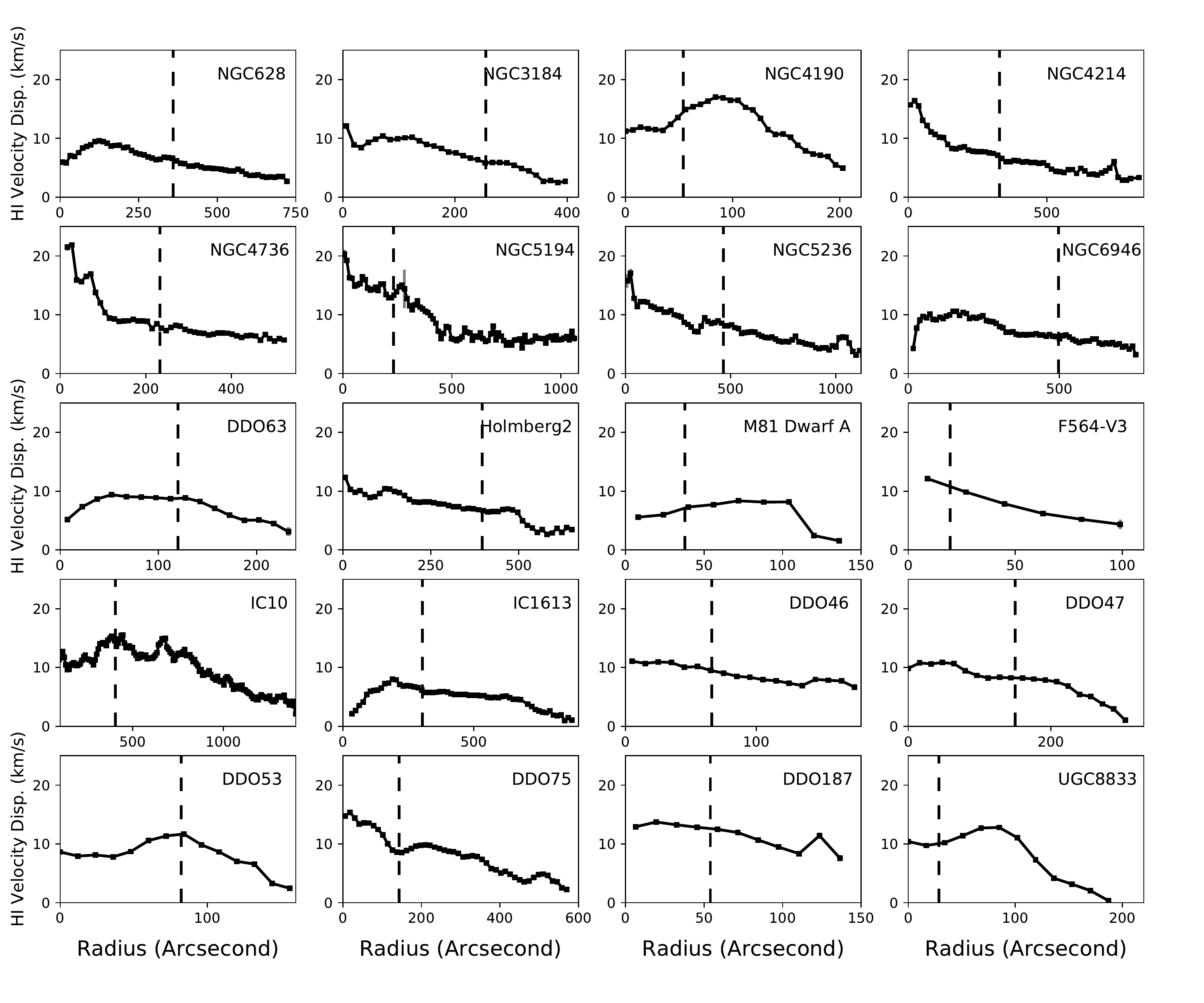} 
    \end{centering}     
    \caption{The \HI~velocity dispersion plotted against the radius of the galaxy. The dashed line is the $R_{25}$ radius.}
    \label{fig:veldisp}
 \end{figure*}

The HI velocity dispersion curves that are used for calculating the galaxy halo oblateness are shown in Figure \ref{fig:veldisp}.

\end{document}